\documentclass[preprint,prd,aps,showpacs,showkeywords,onecolumn,english]{revtex4}
\usepackage{amsmath,amssymb} 
\usepackage[dvips]{graphicx} 
\usepackage{graphics}
\usepackage{graphicx}
\usepackage{epsfig}

\newcommand\nn{\nonumber}
\newcommand\ba{\begin{eqnarray}}
\newcommand\ea{\end{eqnarray}}

\begin{document}

\title{Decays of  $\tau \rightarrow \rho(770) (\rho'(1450)) \nu_\tau$ and  $\tau \rightarrow K^{*}(892) (K^{*'}(1410)) \nu_\tau$
in the extended Nambu - Jona- Lasinio model}

\author{A. I. Ahmadov$^{a,b}$ \footnote{E-mail: ahmadov@theor.jinr.ru},Yu. L. Kalinovsky$^c$ \footnote{E-mail: kalinov@jinr.ru},
M. K. Volkov$^a$,\footnote{E-mail: volkov@theor.jinr.ru}}
\affiliation{$^{a}$ Bogoliubov Laboratory of Theoretical Physics,
JINR, Dubna, 141980 Russia}
\affiliation{$^{b}$ Institute of Physics, Azerbaijan
National Academy of Sciences, H.Javid ave. 131, AZ-1143 Baku, Azerbaijan}
\affiliation{$^{c}$ Laboratory of Information Technologies,
JINR, Dubna, 141980 Russia}

\begin{abstract}
In the extended Nambu - Jona - Lasinio model the decay widths
$\tau \rightarrow \rho(770) (\rho'(1450)) \nu_\tau$ and
$\tau \rightarrow K^{*}(892) (K^{*'}(1410)) \nu_\tau$ are studied in the quark one -loop
approximation.
Our estimations of the decay widths
$\tau \rightarrow K^{*}(892) (K^{*'}(1410)) \nu_\tau$
are in satisfactory agreement  with  experimantal data.
In the paper, the decay widths $\tau \rightarrow \rho(770) (\rho'(1450)) \nu_\tau$  are also calculated.

\vspace*{1.5cm}


\end{abstract}

\maketitle

\section{Introduction}

Recently, in the framework of the  extended Nambu - Jona - Lasinio
(NJL) model  \cite{1997,19971,1998,2000}  a number of processes connected with
the creation of  mesons in $\tau$ decays  and in the reaction of $e^+e^- \rightarrow hh$ at low energy  were successfully described.

Such processes are $\tau \rightarrow \pi^- \pi^0 \nu_\tau$ \cite{a4}, $\tau \rightarrow \eta (\eta') \pi^-  \nu_\tau$ \cite{a1},
$\tau \rightarrow \eta(550)  (\eta'(950)) 2 \pi \nu_\tau$ \cite{a2},
$\tau \rightarrow \pi^- \omega \nu_\tau$ \cite{a3}.
In these reactions it is necessary the take into account in the intermediate states
both the ground state $\rho(770)$ and  the first radial excited state $\rho'(1450)$.

A similar mechanism can be used for the description of the reactions
$e^+e^- \rightarrow hh$ at low energy. Here  the intermediate
$\rho^0$, $\omega$, $\phi$ mesons and their first radial excited states are used.
These processes are
$e^+e^- \rightarrow \pi^0 (\pi^{0'}) \gamma$  \cite{a5},
 $e^+e^- \rightarrow (\eta(550), \eta'(950), \eta(1295), \eta(1475)) \gamma$ \cite{a8},
 $e^+e^- \rightarrow \pi \pi (\pi' (1300))$  \cite{a7},
$e^+e^- \rightarrow \pi^0 \omega$ \cite{a6},
$e^+e^- \rightarrow \pi^0 \rho^0$ \cite{a9},
$e^+e^- \rightarrow \eta(550) (\eta'(950)) 2 \pi$ \cite{a2}.

Naturally it is interesting to describe the decays
$\tau \rightarrow \rho(770) (\rho'(1450)) \nu_\tau$ which are the basis of
the above-mentioned processes. This paper is devoted to the solution of this problem.
Also, it is very interesting to consider the decays
$\tau \rightarrow K^{*}(892) (K^{*'}(1410)) \nu_\tau$ as there are
reliable experimental data for them \cite{pdg}.
 It is shown that our results  obtained in the framework  of the extended NJL model are in satisfactory agreement
 with  these  experimental data.

\section{Lagrangian of the quark - meson interactions in the extended
Nambu - Jona -Lasinio model}

The Lagrangian of the quark - vector meson interactions in the extended
Nambu - Jona -Lasinio model  has the following form:
\begin{eqnarray}
\Delta\mathcal{L}^{int} &=& \bar{q}(k') \bigl[ i\hat{\partial} - {m} +
A_{\rho} \lambda_3 \gamma_\mu \rho_\mu  (p)
-  A_{\rho'}\lambda_3\gamma_\mu \rho'_\mu  (p)   \nonumber \\
&& + A_{K^{\star}}\lambda_{\pm} \gamma_\mu K^{\star}_\mu  (p)  -
A_{K^{\star '}}\lambda_{\pm} \gamma_\mu K^{\star'}_\mu  (p) \bigr] q(k),
\label{L1}
\end{eqnarray}
where $\hat{\partial} = \gamma_\mu \partial_\mu $,
${m} = \mbox{diag} ( m_u, m_d, m_s )$, $m_u=m_d=280$ MeV, $m_s = 405$ MeV,
$q$ and $\bar{q}$ are the quark fields, $\rho_\mu (\rho'_\mu)$ and $K^*_\mu (K^{*'}_\mu)$ are
the vector meson fields in the ground (excited) state
\ba
A_{\rho} &=& g_{\rho_1} \frac{\sin(\beta+\beta_0)}{\sin(2\beta_0)}+g_{\rho_2} f(k_\bot^2)\frac{\sin(\beta-\beta_0)}{\sin(2\beta_0)}, \nn \\
A_{\rho'} &=& g_{\rho_1} \frac{\cos(\beta+\beta_0)}{\sin(2\beta_0)}+g_{\rho_2} f(k_\bot^2)\frac{\cos(\beta-\beta_0)}{\sin(2\beta_0)}, \nn \\
A_{K^{\star}} &=&g_{K^*} \frac{\cos(\theta+\theta_0)}{\sin(2\theta_0)} + g_{K^{*'}} f(k_\bot^2)\frac{\cos(\theta-\theta_0)}{\sin(2\theta_0)}, \nn \\
A_{K^{\star'}} &=&- g_{K^*} \frac{\sin(\theta+\theta_0)}{\sin(2\theta_0)} - g_{K^{*'}} f(k_\bot^2)\frac{\sin(\theta-\theta_0)}{\sin(2\theta_0)},
\ea
\ba
\lambda_3 &=& \left(\begin{array}{ccc} 1&0&0 \\0&-1&0 \\0&0&0 \end{array} \right), \qquad \nn
\lambda_+ = \sqrt 2 \left (\begin{array}{ccc} 0&0&1 \\ 0&0&0 \\0&0&0 \end{array} \right), \qquad \nn
\lambda_- = \sqrt 2 \left (\begin{array}{ccc} 0&0&0 \\ 0&0&0 \\1&0&0 \end{array} \right). \nn
\ea
The values of the angles $\beta=79.85^\circ$ and $\beta_0=61.44^\circ$ are taken from \cite{19971},
and $\theta=84.7^\circ$, $\theta_0=59.14^\circ$ \cite{2000} are the mixing angles for the ground and
first radially excited states of  mesons, respectively.

Radially excited states are described in the extended NJL model using the following form factors $f(k_\bot^2)$  in the quark-meson interaction:
\ba
f(k_\bot^2)=(1-d |k_\bot^2|)\Theta (\Lambda_3^2 -|k_\bot^2|), \nn \\
k_\bot =k-\frac{(kp)p}{p^2}, \,\,\,\,\,d=-1.784~\mathrm{GeV}^{-2},
\ea
 where $k$ and $p$ are the quark and meson momenta, respectively, and the cut-off parameter $\Lambda_3 = 1.03~$ GeV.
 The quark-meson coupling constants are
\ba
g_{\rho_2} = \left(\frac23 I_2^{f^2}(m_u, m_d)\right)^{-1/2}=9.87, \qquad
g_{\rho_1} = \left(\frac23 I_2^{(0)}(m_u, m_d)\right)^{-1/2}=6.14, \nn \\
g_{K^{*'}} = \left(\frac23 I_2^{(f^2)}(m_u, m_s)\right)^{-1/2}= 10.86,  \qquad
g_{K^*} = \left(\frac23 I_2^{(0)}(m_u, m_s)\right)^{-1/2}=6.77,
\ea
where the integrals $I_m^{f^n}$ read
\ba
I^{f^{n}}_m(m_q) = -i \frac{N_c}{(2\pi)^4} \int \mbox{d}^4 k\frac{(f_q({k_\bot}^2))^n}{(m_q^2-k^2)^m}\Theta(\Lambda^2_3 -  k_\bot^2),
\ea
where $N_c =3$ is the number of color.

\section{Amplitudes of  the  decays $\tau \to V (V') \nu_{\tau}$ in the extended NJL model}

\begin{figure}[!htb]
       \centering
       \includegraphics[width=0.5\linewidth]{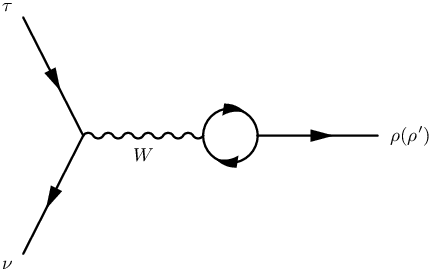}
       \caption{The Feynman diagram for the decay $\tau \rightarrow \rho (\rho') \nu_{\tau}$.}
       \label{decay1}
\end{figure}

The Feynman diagram
for the decay $\tau \to \rho (\rho') \nu_{\tau}$ is shown  on Fig. \ref{decay1}.
The  amplitude of this decay has the form

%
\ba
A_{\tau \to \rho (\rho')\nu_{\tau}}=\frac{G_F}{\sqrt{2}} \cdot \bar{u}_{\nu_\tau}\gamma_{\alpha} u_{\tau} \cdot g_{\alpha \mu} \cdot |V_{ud}|\frac{g_{\rho}}{2}
\int\frac{d^4k}{(2\pi)^4}\frac{tr\,\, [ \gamma_{\mu}  ( (\hat {k}+\hat {p})+m_u)\gamma_{\nu} (\hat {k} +m_u]e^{\nu}_{\rho}(p_{\rho})}{(k^2-m_u^2)((k+p)^2-m_u^2)}.
\label{A1}
\ea
Here $p$ is the $\rho$ -  meson momentum, $G_F = 1.16637 \cdot 10^{-11} MeV^{-2}$ - is the Fermi constant,
$k$  is  the quark momentum, $m_u$ - is the $u$ - quark mass, and
$|V_{ud}|$=0.97428 is the Cabibbo - Kobayashi - Maskawa mixing  angle.

The square of the amplitude takes the form
\ba
|M|^2 = 4 m_{\tau} m_{\rho (\rho')}^2 E_{\nu} |V_{ud}|^2 \frac{G_F^2}{2}\frac{1}{g_{\rho (\rho')}^2}\biggl[2E_{\rho (\rho')}^2 +
m_{\rho (\rho')}^2 -2E_{\rho (\rho')}\sqrt{E_{\rho (\rho')}^2 -m_{\rho (\rho')}^2} \biggr]
\label{M2}
\ea
The decay width for the process is
\ba
\Gamma (\tau \to \rho (\rho') \nu_{\tau}) = \frac{|M|^2}{2 \cdot 2 m_{\tau}}\Phi,
\label{Gamma}
\ea
where  $\Phi$  is the phase volume:  
\ba
\Phi = \frac{E_{\nu}}{4\pi m_{\tau}},
\ea
and  $E_{\nu}$  and  $E_{\rho}$ are determined  as
\ba
E_{\nu} = \frac{m_{\tau}^2 - m_{\rho (\rho')}^2}{2 m_{\tau}}, \qquad
E_{\rho} = \frac{m_{\tau}^2 + m_{\rho (\rho')}^2}{2 m_{\tau}},
\ea
We also use $(p_{\nu}p_{\tau}) = m_{\tau} E_{\nu}$, \quad
$(p_{\tau}p_{\rho (\rho')}) = m_{\tau} E_{\rho (\rho')}$.

\begin{figure}[!htb]
       \centering
       \includegraphics[width=0.5\linewidth]{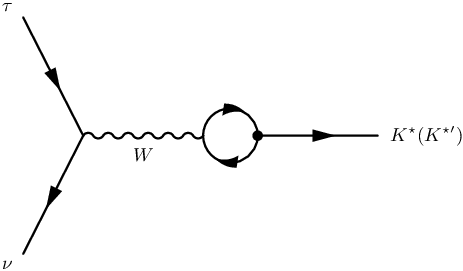}
       \caption{The Feynman diagram for the decay  $\tau \rightarrow K^{\star} (K^{\star'}) \nu_{\tau}$.}
       \label{decay2}
\end{figure}

The Feynman diagram
for the decay  $\tau \to K^{\star} (K^{\star'}) \nu_{\tau}$ is shown  on Fig. \ref{decay2} and
the amplitude can be written as

\ba
A_{\tau \to K^{\star} (K^{\star'}) \nu_{\tau}}=\frac{G_F}{\sqrt{2}} \cdot \bar{u}_{\nu_\tau}\gamma_{\alpha} u_{\tau} \cdot g_{\alpha \mu} \cdot |V_{us}|\frac{g_{K^{\star}}}{2}
\int\frac{d^4k}{(2\pi)^4}\frac{tr\,\, [ \gamma_{\mu}  ( (\hat {k}+\hat {p})+m_u)\gamma_{\nu} (\hat {k} +m_s]e^{\nu}_{\rho}(p_{\rho})}{(k^2-m_s^2)((k+p)^2-m_u^2)},
\label{A2}
\ea
where  the $m_s$ is the $s$- quark mass, and
$|V_{us}|$=0.2252 is the Cabibbo - Kobayashi - Maskawa mixing  angle.

Using formula \eqref{Gamma} for the decay width $\tau \to \rho (\rho') \nu_{\tau}$, and the following numerical results we obtain
\ba
\Gamma_{\tau \to \rho \nu_{\tau}}^{theor} =2.98 \cdot 10^{-11} \,\,\,MeV,
\ea
and
\ba
\Gamma_{\tau \to \rho' \nu_{\tau}}^{theor} =3.306 \cdot 10^{-12} \,\,\,MeV.
\ea
The square of the amplitude has an analogous form \eqref{M2} with the replacement
$\rho (\rho') \to K^\ast (K^{\ast '})$ and $|V_{ud}| \to |V_{us}|$.

The numerical result
\ba
\Gamma_{\tau \to K^\ast \nu_{\tau}}^{theor} =2.60 \cdot 10^{-11} \,\,\,MeV,
\ea
and
\ba
\Gamma_{\tau \to K^{\ast '} \nu_{\tau}}^{theor} =5.15 \cdot 10^{-12} \,\,\,MeV.
\ea
The experimental data are $\Gamma_{\tau \to K^\ast \nu_{\tau}}^{exp}=(2.72 \pm 0.15) \cdot 10^{-11}$ MeV \cite{Olive}, and
$\Gamma_{\tau \to K^{\ast '} \nu_{\tau}}^{exp} = 3.4 \left(\begin {array}{c}+3.178 \\-2.27 \end{array} \right) \cdot 10^{-12}$ \,\,\,MeV \cite{Olive} .

\section{Discussions and Conclusion}
The presented here calculations demonsrate that the extended NJL model allows us to
describe the decays
$\tau \rightarrow K^{*}(892) (K^{*'}(1410)) \nu_\tau$ in satistactory agreement
with experimental data. Let us  emphasize that these results were obtained
without  using any additional arbitrary parameters.
The corresponding estimations  for the decay width
$\tau \rightarrow \rho(770) (\rho'(1450)) \nu_\tau$ are also obtained.

The calculations of the amplitude of the decay $\tau \rightarrow V \nu_\tau$,
where $V$ is the vector meson field, in the one - quark loop approximation
in the  NJL  model  take  the gradient invariant form
$g_{\mu \nu} p^2 - p_\mu p_\nu$.
Let us note that both terms for the decay width $\tau \rightarrow V \nu_\tau$
in the expression (\ref{A2}) play an important role. Indeed, if we use for the description of
both decays  $\tau \rightarrow V (V')\nu_\tau$, where $V'$ is the first radially excited state,
the term  $g_{\mu\nu} p^2$, as in \cite{Okun}, then for the decay
$\tau \rightarrow V' \nu_\tau$ a wrong result will be obtained.
On the other hand, it interesting that for  all  more complicated processes
discussed  in the Introduction, where  vector mesons are the intermediate states,
the term $p_\mu p_\nu$ automatically  gives  zero after multiplication by the vertex
describing the vector meson transition to the final product of the corresponding  decay.
As a result, the diagram containing the term $g_{\mu\nu} p^2$ together with the contact
diagram, where $W$ directly goes to the final product through the quark loop,
leads to the vector dominant model. It explains the success of
the vector dominant model for the description of different $\tau$ decays
\cite{a9,pdg,1986,a15,volkov10,a16,a11,a12,a13,a14}.
However, in these phenomenological models,
for a satisfactory description of experimental data
it is necessary to use a set of  arbitrary parameters.
The extended NJL model  allows us to describe the $\tau$ decays and $e^+e^-$ processes
at low energy without  introduction of any additional arbitrary parameters.
Using our  model in future works  we are going to consider
more complicated $\tau$ decays, in particular, decays with
participation of strange particles.

\section{Acknowledgements}

We are gratefull to A. B. Arbuzov for the useful discussions. This work has been supported in part by the RFBR
grant no. 13-01-00060a (Yu. L. K).

\end{document}